\documentclass[10pt]{article}

\usepackage[OE]{express2}
\usepackage{braket}
\usepackage{amsmath}
\usepackage{amsfonts}
\usepackage{amssymb}
\usepackage{graphicx}
\usepackage{color}

\newcommand{\ReviewedText}[1]{{#1}}

\begin{document}
\title{Laguerre-Gauss beam generation in IR and UV by subwavelength surface-relief gratings}

\author{Larissa Vertchenko,\authormark{1} Evgeniy Shkondin,\authormark{2} Radu Malureanu,\authormark{3} and Carlos Monken\authormark{1,*}}

\address{\authormark{1}Departamento de F\'{\i}sica, Universidade Federal de Minas Gerais, Caixa Postal 702, Belo Horizonte, MG 30161-970, Brazil\\
\authormark{2}Department of Photonics Engineering, Technical University of Denmark, \O rsteds Plads 343 DK-2800 Kgs. Lyngby, Denmark and Danish National Center for Micro and Nanofabrication (DANCHIP), DK-2800 Kongens Lyngby, Denmark\\
\authormark{3}Department of Photonics Engineering, Technical University of Denmark, \O rsteds Plads 343 DK-2800 Kgs. Lyngby, Denmark}

\email{\authormark{*}monken@fisica.ufmg.br} 

\begin{abstract}
The angular momentum of light can be described by the states of spin angular momentum, associated with polarization, and orbital angular momentum, related to the helical structure of the wave front. Laguerre-Gaussian beams carry orbital angular momentum and their generation can be done by using an optical device known as q-plate. However, due to the usage of liquid crystals, these components may be restricted to operate in specific wavelengths and low power sources. Here we present the fabrication and characterization of q-plates made without liquid crystals, using processes of electron beam lithography, atomic layer deposition and dry etch techniques. We exploit the phenomenon of form birefringence  to give rise to the spin-to-orbital angular momentum conversion. We demonstrate that these plates can generate beams with high quality for the UV and IR range, allowing them to interact with high power laser sources or inside laser cavities.
\end{abstract}

\ocis{(050.4865) Optical vortices; (140.3300) Laser beam shaping; (050.6624) Subwavelength structures; (050.2555) Form birefringence.} 


\section{Introduction}

Q-plates are well known devices that couple spatial and polarization degrees of freedom of optical beams \cite{Marrucci:2006ga}. Due to this coupling, it is possible to exchange electromagnetic angular momentum between circular polarization modes (spin angular momentum) and helical wavefront modes (orbital angular momentum) \cite{2000ConPh..41..275P,doi:10.1117/12.722305}. The orbital angular momentum (OAM) eigenstates \cite{VanEnk1994}, such as the Laguerre-Gauss (LG) transverse modes \cite{siegman}, are characterized in the cylindrical coordinate system by the azimuthal phase factor $\exp(il\phi)$, where $l$ is an integer and $\phi$ the azimuthal angle.  These modes have many applications including fiber mode selection \cite{gregg2015q}, optical tweezers and micromanipulation \cite{yao2011orbital}. In the domain of quantum optics, LG  beams carry an OAM of $l\hbar$ per photon \cite{Padgett:2014} and define an  infinite basis in Hilbert space that can be represented, along with the spin angular momentum states, in a higher-order Poincar\'e sphere \cite{Naidoo:2016iv}.  

\begin{figure}[hbt]
\centerline{\includegraphics{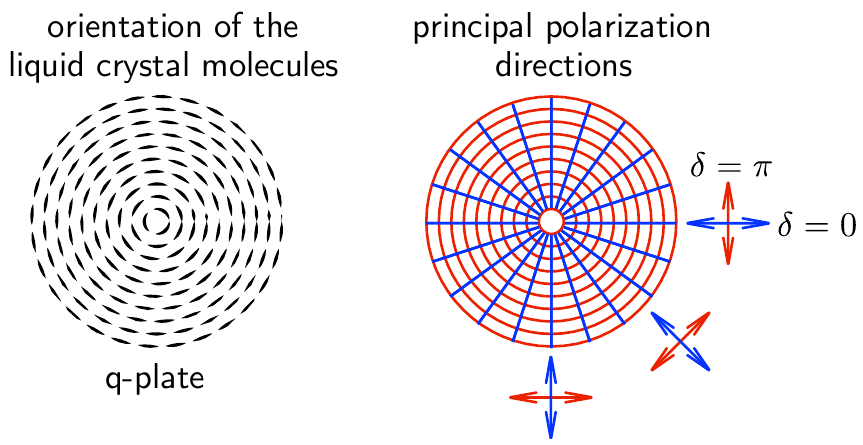}}
\caption{Example of a liquid crystal q-plate whose molecules are azimuthally oriented. The phase shifts are $\delta= 0$ for linear polarization along the radial direction and $\delta=\pi$ for linear polarization along the azimuthal direction.}
\label{fig:pol}
\end{figure}

In order to generate an helical phase front, which carries OAM, one must introduce a convenient polarization and position dependent phase shift $\delta(\sigma,\mathbf{r})$ on the beam cross section. Usually, this task is realized in a q-plate by a thin layer of liquid crystal whose molecules have their axes oriented along a certain pattern engraved on a glass surface \cite{Marrucci:2008ed}. For example, when a circularly  polarized Gaussian beam passes through a q-plate with \ReviewedText{$\delta= 0$  ($\delta=\pi$)} for linear polarization along the radial (azimuthal) direction, its polarization handedness is flipped and it acquires an OAM of $\pm 2\hbar$. This situation is illustrated in Fig.  ~\ref{fig:pol}. \ReviewedText{The right value of $\delta$ is achieved by tuning the q-plate thermally \cite{Karimi:2009ge}, electrically \cite{2011OptL...36..719B,Slussarenko:2011dd}, or by photoalignment \cite{Ji:2016}.} 

\ReviewedText{Although the use of a liquid-crystal allows for relatively simple fabrication techniques \cite{Marrucci:2006kd}, its absorption spectrum may set limits to the wavelength and power ranges that can be handled by the q-plate.  Several alternatives to liquid crystals in q-plate fabrication have been demonstrated. One interesting and commercially available technique relies on the phenomenon of form birefringence of subwavelength gratings produced by direct femtosecond laser writing in fused silica plates \cite{Beresna:2011}. This method, though simple and elegant, is limited by the laser writing resolution, resulting in devices intended to work with wavelengths in the visible and near infrared. Dielectric, Si, and metal-based metasurfaces made of dense arrays of nanopillars, holes or optical antennas have been demonstrated to behave as q-plates, but also limited to wavelengths larger than approximately 500\,nm \cite{Ghadyani:2011,Li:2013,Lin:2014,Bouchard:2014,Karimi:2014,Hakobyan:2015,Arbabi:2015,Mehmood:2016}. 

Another way to produce q-plates with form birefringence is the fabrication of subwavelength surface relief gratings by photolithography \cite{Biener:7VTdkd_M,Niv:2005cl,2006ApOpt..45.1777T}. However, due the relatively low resolution achieved with this technique, the q-plates are in general limited to wavelengths of the order of 10\,$\mu$m. More recently, electron beam lithography (EBL)  combined with atomic layer deposition (ALD) of TiO$_2$ has been used to achieve unprecedented resolutions, resulting in q-plates capable of operating in shorter wavelengths \cite{Devlin:2017}. The short wavelength limit is, in principle, defined by the $\sim$\,20\,nm resolution of EBL.

In this paper we demonstrate a TiO$_2$-based q-plate fabricated with EBL combined with ALD, capable of operating in a large wavelength range, from IR to UV. Differently from Ref. \cite{Devlin:2017}, the pattern formed by the structures was chosen to be concentric circles of sub-wavelength grooves with the appropriate depth. This patern is able to generate LG beams of $l=\pm 2$ as explained below. Due to the higher density of this concentric circles structure, compared to the structure of nanofins used in Ref. \cite{Devlin:2017}, higher conversion efficiencies are, in principle, attainable. We show that the q-plates fabricated with this technique are capable of handling high peak powers without any damage or distortions.  We also investigated the interaction of the LG beams generated by q-plates with a nonlinear crystal, resulting in second harmonic generation (SHG), and how it affects the order and, consequently, the wave front of the incident modes. The q-plates were characterized using an asymmetric Mach-Zehnder interferometer and a CCD camera, showing the generation of high quality LG modes in both 810 and 405\,nm. Our results are of particular interest in the field of quantum optics, since we produce light beams carrying OAM with enough power in the UV to generate entangled photon pairs through spontaneous parametric downconversion.}

\ReviewedText{Starting from a pulsed laser beam with an average power of 3.6\,W,  140\,ns pulses centered at $\lambda=810$\,nm at 80\,MHz repetition rate (resulting in a peak power of  0.32\,MW) and its second harmonic, we produced pulsed LG modes with $l=2$, 4 and 6 centered at $\lambda=405$\,nm.}

\section{Form birefingence}
\begin{figure}
\centerline{\includegraphics{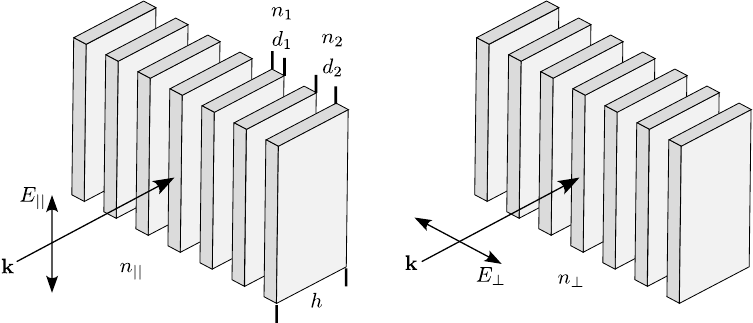}}
\caption{Form birefringence: a material of refractive index $n_1$ is arranged in a periodic nano structure with a filling factor $f$, and surrounded by a medium of refractive index $n_2$. Light linearly polarized in the directions parallel ($E_{||}$) and normal ($E_{\perp}$) to the interfaces will propagate with different velocities, corresponding to two different refractive indices $n_{||}$ and $n_\perp$.}
\label{fig:structure}
\end{figure}

Form birefringence may arise when sub-wavelength structures of different refractive indices are arranged with some order. According to the effective medium theory (EMT) \cite{born1999principles}, a periodic array of parallel slabs of thickness $d_1$ and refractive index $n_1$ separated by a gap $d_2$, embedded in a medium of refractive index $n_2$, as depicted in  Fig.  ~\ref{fig:structure}, will present a birefringence given by
\begin{equation}
\Delta n = n_{||}-n_\perp=\sqrt{fn_1^2+(1-f)n_2^2} - \sqrt{\frac{f}{n_1^2}+\frac{1-f}{n_2^2}},
\end{equation}
where $n_{||}$ and $n_\perp$ are the effective refractive indices corresponding to linear polarizations along the directions parallel and perpendicular to the slabs, respectively, and $f=d_1/(d_1+d_2)$ is the filling factor.

Considering normal incidence and $n_2=1$ (air), diffraction effects will be absent when the structures have a maximum spatial period equivalent to $d_1+d_2=\Lambda = {\lambda/n_1}$ \cite{Raguin:1993}. To achieve the desired phase retardation $\delta$ for our q-plates, the height $h$ of the structures must be adjusted  according to   $h=\lambda\delta/(2\pi\Delta n)$. 
 
\section{Design and fabrication}
\begin{figure}[hbt]
\centerline{\includegraphics{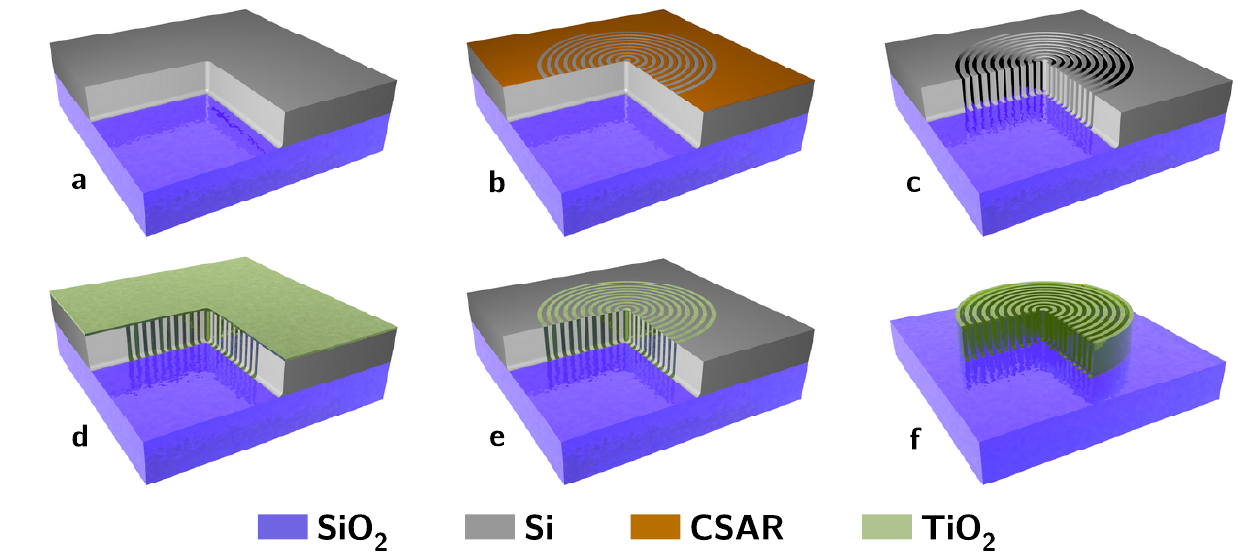}}
\caption{Scheme of fabrication flow. (a) Si deposited on $\textnormal{SiO}_2$ by LPCVD process (only top part considered, sub sequential removal of the backside is not included). (b) E-beam lithography. (c) Advance silicon etching (ASE). (d) ALD of amorphous $\textnormal{TiO}_2$. (e) Removal of  $\textnormal{TiO}_2$ by IBE (only top part considered) $\textnormal{TiO}_2$ structures. (f) Isotropic $\text{SF}_6$ based RIE process and final structure. } \label{fig:process}
\end{figure}

\ReviewedText{The optical elements presented in this article were intended to work at the wavelengths of 810\,nm and its second harmonic (405\,nm) from a pulsed laser source.} The material chosen to form the structures was titanium dioxide ($\textnormal{TiO}_2$), which exhibits a refractive index of 2.52 at $\lambda=810$\,nm and 2.97 at $\lambda=405$\,nm.  In order to extinguish diffraction orders at 810\,nm, the spatial period calculated was $\Lambda= 322$\,nm, which means that for $f=0.5$ the trenches would have a width of 161\,nm. \ReviewedText{The height of the structure was chosen to be 300\,nm. With these dimensions, the structure was expected to have approximately the conversion efficiency for both wavelengths (see section 4), although  some diffraction loss was expected at 405\,nm.} All the work has been carried out in class 100 cleanroom facility. The main steps of the fabrication process are shown in Fig.  \ref{fig:process}.

\ReviewedText{A 500\,$\mu$m-thick wafer of silica} ($\textnormal{SiO}_2$) goes through RCA clean and low pressure chemical vapor deposition (LPCVD) (furnace from
Tempress) based on SiH4 (silane) at 560$^{\circ}$C to form a layer of 680\,nm of amorphous silicon (Si) [Fig.   3(a)]. The back side of deposited Si was etched using sulphur hexafluoride ($\text{SF}_6$). In order to remove residues from the etching process it was performed oxygen plasma cleaning and RCA cleaning. A deposition of 150\,nm of the resist CSAR was done followed by exposure to Electron Beam Lithography (EBL) (JEOL JBX-9500 Electron-beam) [Fig.  3(b)] generating a mask of Si with concentric ring patterns. After development, the wafer was submitted to advanced silicon etch (ASE) [Fig.  3(c)]. 

To form the trenches of the structures, a thin film of $\textnormal{TiO}_2$ was deposited using the ALD technique in a hot-wall system (Picosun R200), working with 2000 cycles at 150$^{\circ}$C [Fig.   3(d)] \cite{shkondin2016fabrication}. The precursors used were titanium tetrachloride ($\text{TiCl}_4$)
and $\text{H}_2\text{O}$ (supplied by Strem Chemicals
Equipment). The process was followed by Ar$^+$ ion beam etching (IBE) on both sides of the wafer to remove excess[Fig.  3(e)]. At the top most $\textnormal{TiO}_2$ layer the physical sputtering of the sample using Ar$^+$ ions was performed in order to get access to Si core. On the backside, the Ar$^+$ ions were used to remove the deposited $\textnormal{TiO}_2$\cite{huang2012atomic}. 

\begin{figure}[h]
\centerline{\includegraphics{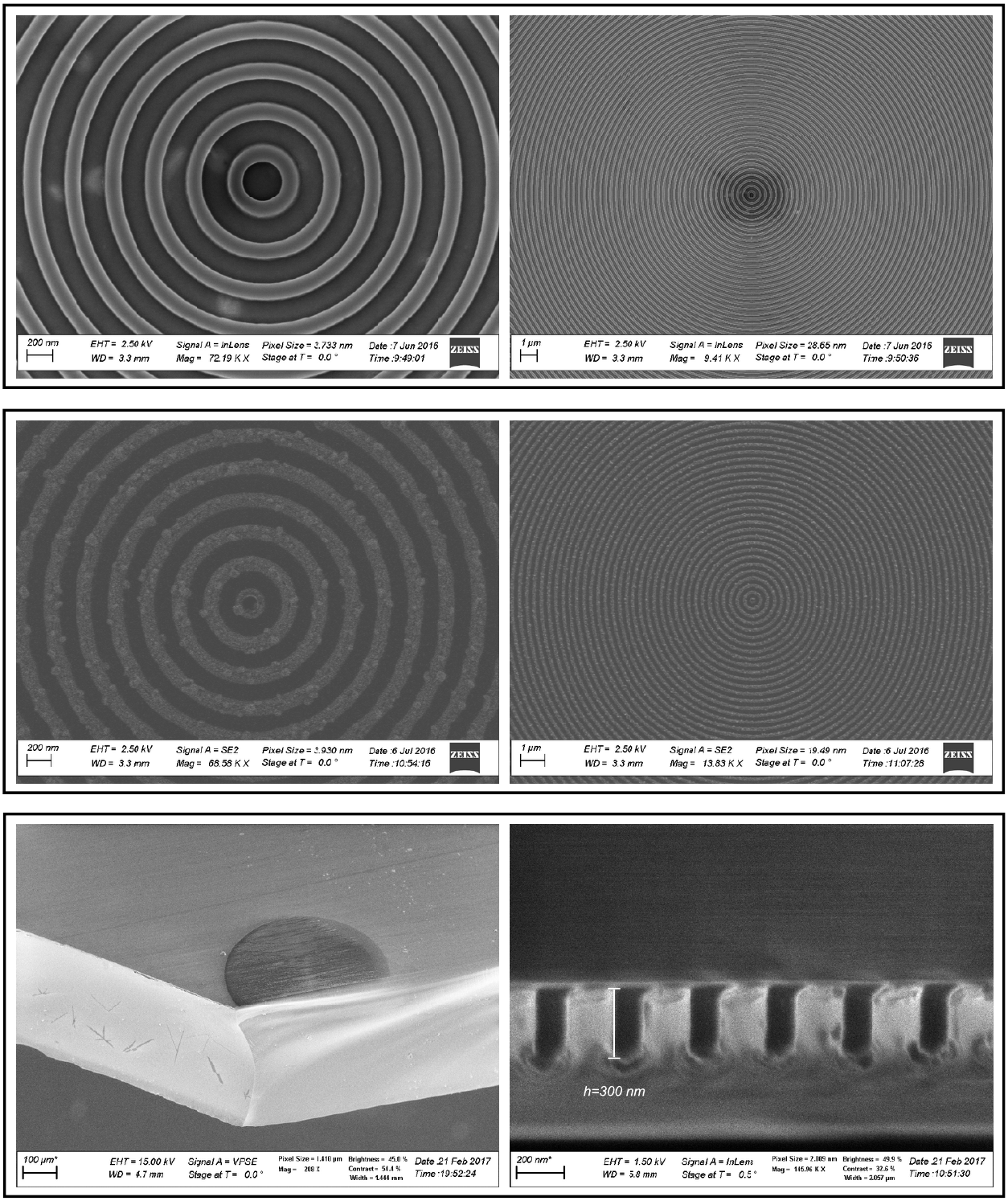}}
\caption{SEM images of the q-plate. Top: Si mask. Middle:  \textnormal{TiO}$_2$ structures. Bottom: Cross section of the \textnormal{TiO}$_2$ structures.} \label{fig:SEM}
\end{figure}

\begin{figure}[hbt]
\centerline{\includegraphics{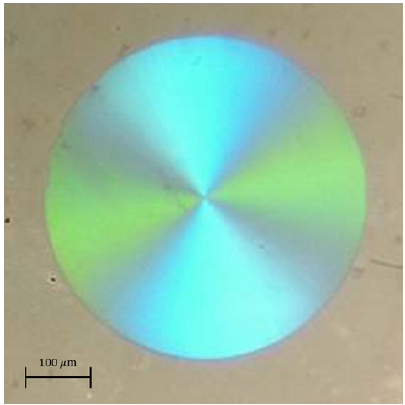}}
\caption{Image of tha q-plate taken from an optical microscope.}\label{fig:plate}
\end{figure}

Finally we performed a reactive ion etch on silicon, leaving only the $\textnormal{TiO}_2$ structures. The final system comprehends a base of $\textnormal{SiO}_2$ with nano-structures of $\textnormal{TiO}_2$ on it [Fig.  3(f)]. Figure \ref{fig:SEM}(a) depicts an image taken from a scanning electron microscope (SEM) of the Si mask and Fig.  \ref{fig:SEM}(b), the final structures of $\textnormal{TiO}_2$.  Figure \ref{fig:plate} shows the image of the system taken from an optical microscope.

\ReviewedText{It should be noted that our process is very similar to the one used for the fabrication of high aspect ratio nickel zone plates for x-ray microscopy \cite{Grenci:2012}.}

\section{Higher order Laguerre-Gauss modes generation}
\begin{figure}[hbt]
\centerline{\includegraphics{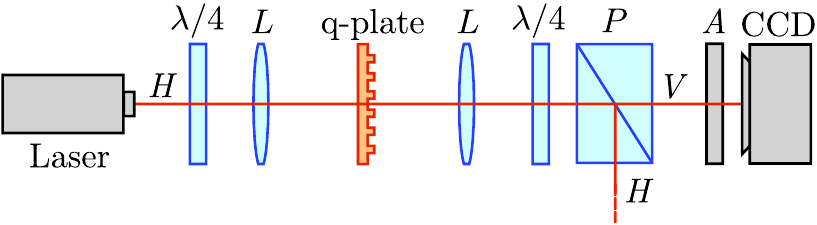}}
\caption{Setup for the generation of a Laguerre-Gauss mode using the fabricated q-plate.}\label{fig:setup1}
\end{figure}

\begin{figure}[hbt]
\centerline{\includegraphics{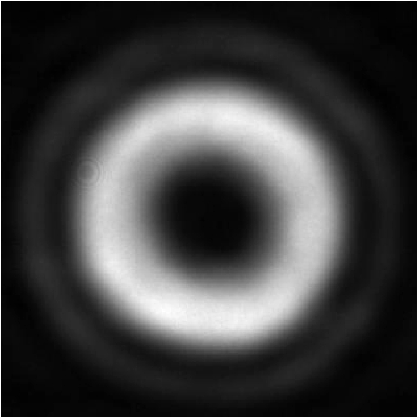}}
\caption{Intensity profile of the Laguerre-Gauss mode of $l=2$ and $\lambda=810$\,nm generated by the q-plate.}
\label{fig:LG2}
\end{figure}

The generation of a LG mode of $l=2$ at $\lambda=810$\,nm through a q-plate is achieved by considering the scheme shown in Fig.  \ref{fig:setup1}. \ReviewedText{A pulsed Gaussian beam of 3.6\,W average power (0.32\,MW peak power) with horizontal polarization and $l=0$} is converted to left circular polarization by a quarter wave plate ($\lambda/4$) and focused by the lens $L$ ($f=12\,$cm) on the q-plate. The outgoing wave is collimated by another lens (same as $L$) and filtered by another quarter wave plate and a polarizer ($P$) that transmits vertical polarization, which corresponds to the converted beam with right circular polarization. \ReviewedText{At the output of the polarizer we observe a Laguerre-Gaussian beam of $l=2$.} The intensity profile of the converted beam captured by a CCD (charge coupled device) camera \ReviewedText{equipped with an attenuator ($A$)} is  shown in Fig.  \ref{fig:LG2}.   

\ReviewedText{Since different kinds of losses such as reflection, absorption and  transmission without conversion (due to the finite value of the grating period) are expected, we did not measure the absolute conversion efficiency. Instead, we measured the relative conversion efficiency defined as $\eta=I_V/(I_V+I_H)$, where $I_V$ and $I_H$ are the intensities transmitted and rejected by the polarizer $P$, as indicated in Fig.   \ref{fig:setup1}. The values measured were 0.45 for $\lambda=810$\,nm and 0.50 for  $\lambda=405$\,nm. Considering that the measured input power was 3.6\,W and the measured transmitted power ($I_V+I_H$)  was 2.9\,W at 810\,nm, we estimate an absolute conversion efficiency (disregarding the losses in the polarizer) of at least 0.36. It should be noted that the q-plates were not optimized to 810\,nm.}

\begin{figure}[hbt]
\centerline{\includegraphics{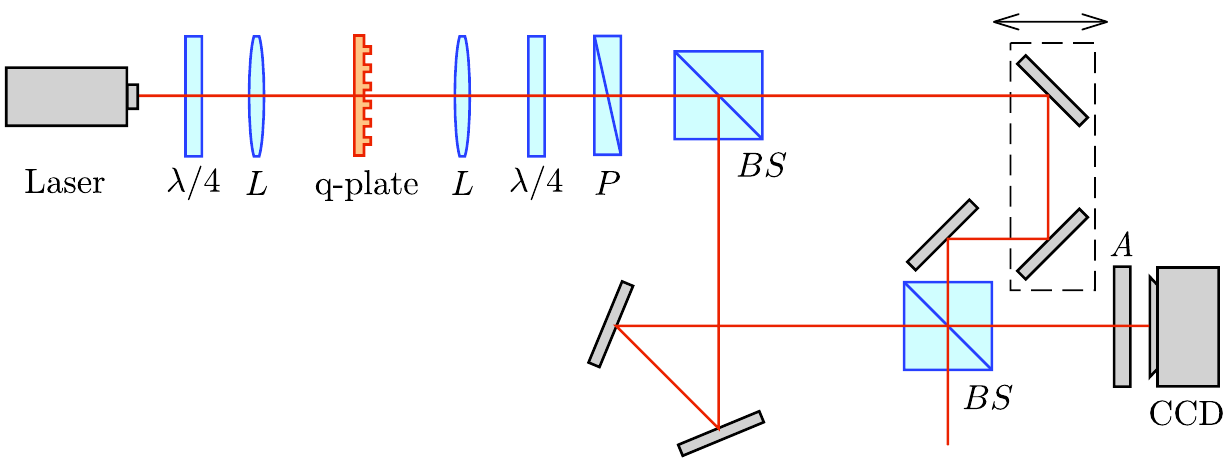}}
\caption{Setup with the asymmetric Mach-Zehnder interferometer.} \label{fig:AMZ}
\end{figure}

\begin{figure}[hbt]
\centerline{\includegraphics{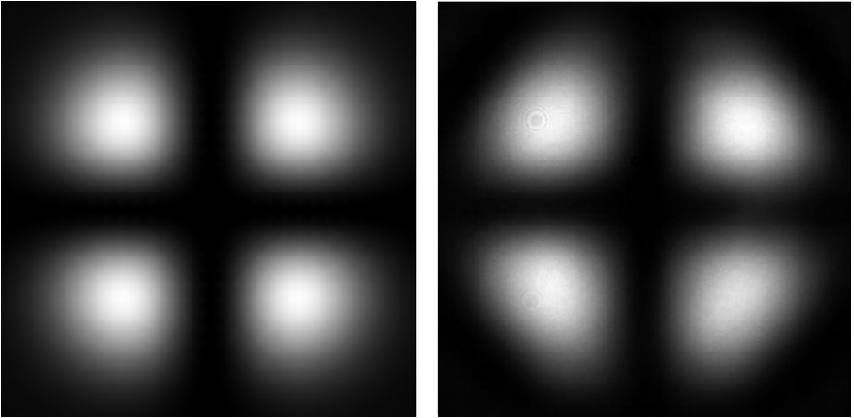}}
\caption{Simulated intensity profile of the interference between LG modes of $l=2$ and $l=-2$ (Left), compared with the one captured by the CCD for $\lambda=810$\,nm (Right).} \label{fig:flower}
\end{figure}

In order to demonstrate that the converted mode is indeed a LG with a helical wave front of $l=2$, it is necessary to perform interferometric measurements. In our experiment we used an asymmetric Mach-Zehnder interferometer, which is able to interfere the transmitted beam with its mirror image, that has an inverted handedness profile due to an additional reflection inside the interferometer. The experimental set-up is illustrated in Fig.  ~\ref{fig:AMZ}, and the simulated intensity profile compared with the one captured by the CCD camera is shown in Fig.  ~\ref{fig:flower}.

\begin{figure}[hbt]
\centerline{\includegraphics{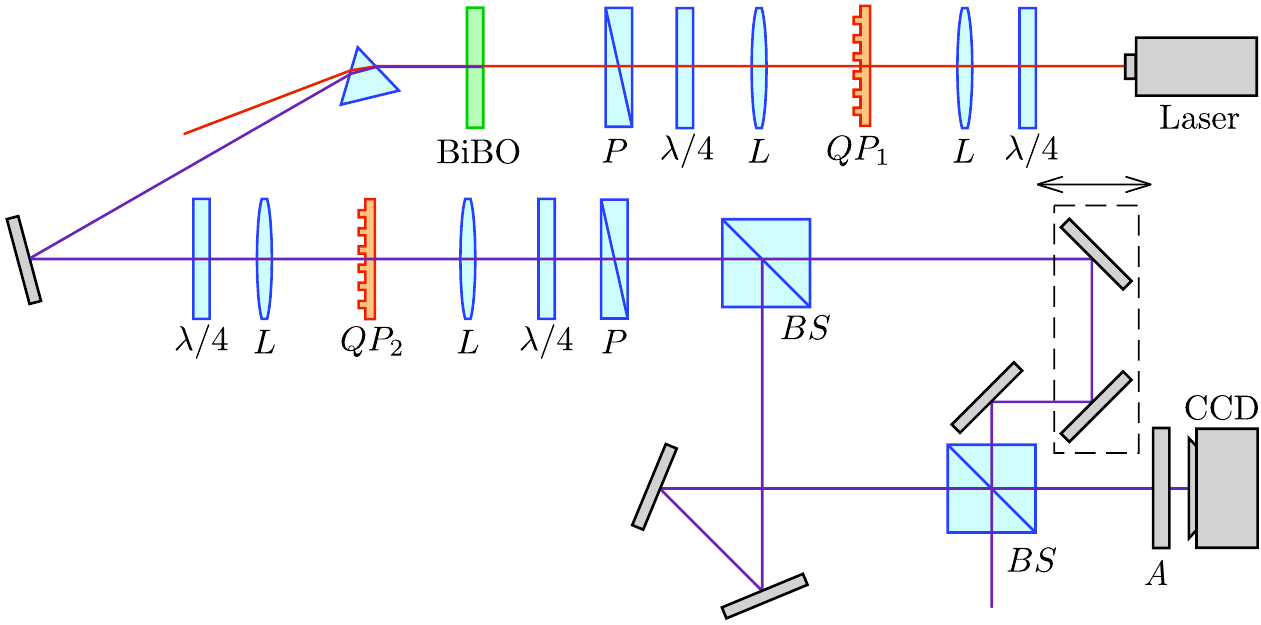}}
\caption{Setup for the generation of Laguerre-Gauss modes of $l=2,\,4,\,6$.} \label{fig:shg}
\end{figure}

\begin{figure}[hbt]
\centerline{\includegraphics{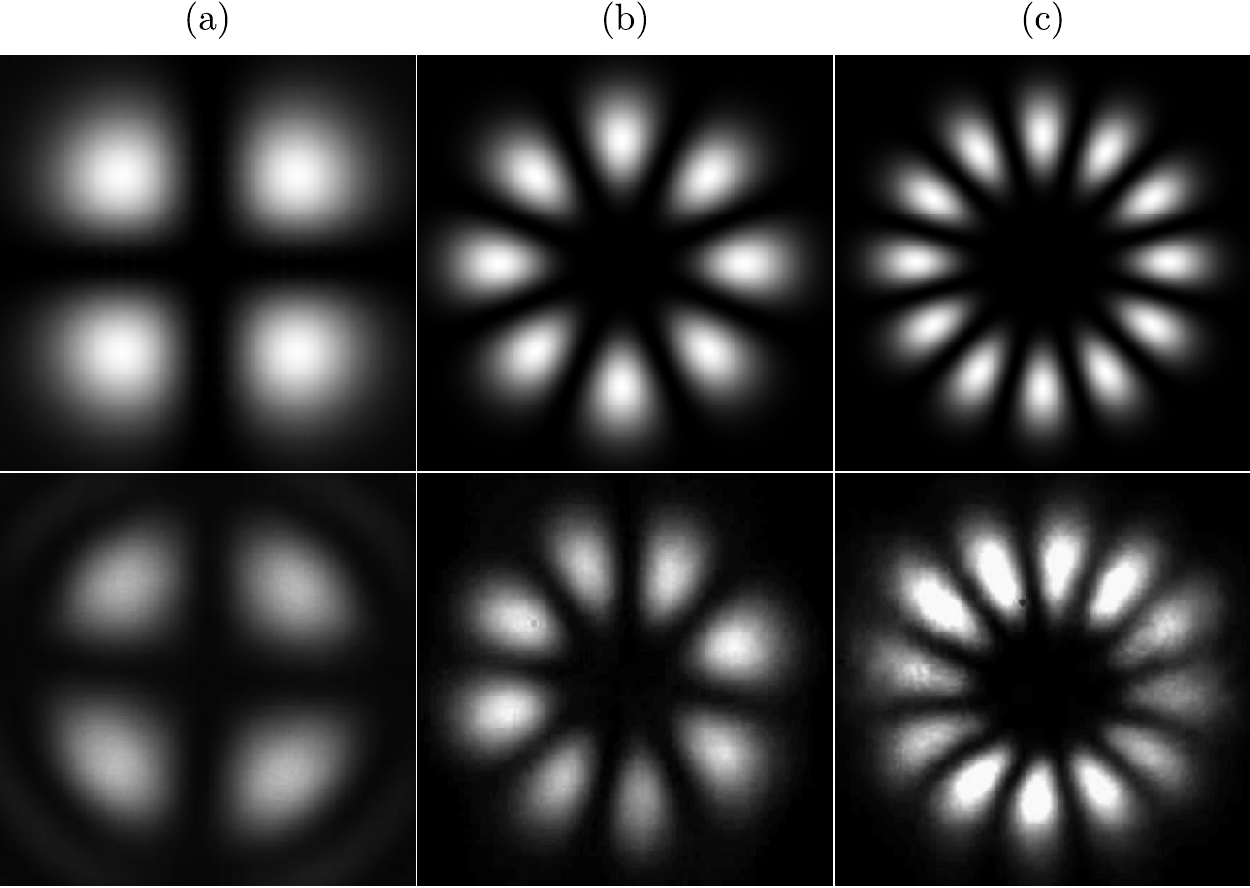}}
\caption{Simulated intensity profiles (top) compared with the ones captured by the CCD camera (bottom) for $l=2\ (a)$, $l=4\ (b)$ and $l=6\ (c)$, $\lambda=405\,$nm, at the output of the asymmetric interferometer.} \label{fig:flowers}
\end{figure}

Having a high quality LG mode generated, we then investigated the possibilities of second-harmonic generation (SHG) and combination of q-plates in order to generate modes with $l=2$, 4 and 6.  It is known that  the orbital angular momentum per photon is doubled in the process of second-harmonic generation (SHG) \cite{courtial1997second}, resulting in a higher order mode of $2l$. The setup used, shown in Fig.  ~\ref{fig:shg}, consists of a incident infrared pulsed beam ($\lambda=810$\,nm) going through a \ReviewedText{1\,mm-long} BiB$_3$O$_6$  nonlinear crystal (BiBO), which produces a frequency-doubled ($\lambda=405$\,nm) beam. The asymmetric interferometer was positioned at the output to verify that the desired value of $l$ was produced. If the q-plate $QP_2$ only is in place, an $l=2$, $\lambda=405$\,nm beam is obtained, just as in the infrared case discussed previously. 
When the q-plate $QP_1$ only is in place, an $l=2$, $\lambda=810$\,nm beam is produced and converted to an $l=4$, $\lambda=405$\,nm  beam by SHG. Finally, when both q-plates $QP_1$ and $QP_2$ are in place, an $l=6$, $\lambda=405$\,nm beam is obtained. 
The intensity profile captured by CCD after the interferometer in all cases is shown in Fig.  ~\ref{fig:flowers}.

\clearpage
\section{Conclusion}
We have demonstrated the generation of LG modes using q-plates exploring the phenomenon of form birefringence. The plates show high optical quality, operating with a high power pulsed laser source. Although we have used the same q-plate for 405\,nm and 810\,nm, they can be fabricated for the specific wavelengths, leading to higher conversion efficiencies.  Our approach creates the possibility to easily fabricate phase plates with different patterns for any wavelength and without the necessity of liquid crystals.  We have also proven the conservation of angular momentum during second-harmonic generation, producing beams with the orbital angular momentum doubled. By combining a set of q-plates and second harmonic generation, we were able to generate LG modes with $l=2$, 4 and 6. The results presented in this paper may find applications in quantum optics and optical communications, due to the fact that we are capable of generating high quality Laguerre-Gauss modes of different orders, with no damage of the plates as a result of the laser power or the wavelength used. This feature is essential to form high dimension alphabet for optical communications and the prospect to insert these plates inside laser cavities. 

\section*{Funding}
Conselho Nacional de Desenvolvimento Cient\'{\i}fico e Tecnol\'{o}gico (CNPq) (307481/2013-1).


\end{document}